\def\Xint#1{\mathchoice
   {\XXint\displaystyle\textstyle{#1}}%
   {\XXint\textstyle\scriptstyle{#1}}%
   {\XXint\scriptstyle\scriptscriptstyle{#1}}%
   {\XXint\scriptscriptstyle\scriptscriptstyle{#1}}%
   \!\int}
\def\XXint#1#2#3{{\setbox0=\hbox{$#1{#2#3}{\int}$}
     \vcenter{\hbox{$#2#3$}}\kern-.5\wd0}}

\def\dashint{\Xint-}

\documentclass[aps,prb,preprint,eqsecnum]{revtex4} 

\newcommand{\re}[1]{(\ref{#1})}

\newcommand{\Z}[1]{\sqrt{(#1-\mu)^2+\Delta^2}}

\newcommand{\up}{\uparrow}

\newcommand{\dn}{\downarrow}

\newcommand {\dis}{\displaystyle}

\newcommand{\beg}{\begin{equation}}
\newcommand{\en}{\end{equation}}

\newcommand{\eps}{\epsilon}
\newcommand{\lam}{\lambda}

\usepackage{graphics}

\begin{document}

\title{Finite Size Corrections for the Pairing Hamiltonian}

\author{Emil~A.~Yuzbashyan}
\author{Alexander~A.~Baytin}
\author{Boris~L.~Altshuler}

\affiliation{
\centerline{Physics Department, Princeton University, Princeton, NJ 08544}
\centerline{NEC Research Institute, 4 Independence Way, Princeton, NJ 08540}
}

\begin{abstract}
We study the effects of superconducting pairing in small metallic grains.
We show that in the limit of large Thouless
conductance one can explicitly determine the low energy spectrum of the problem as an expansion
in the inverse number of electrons on the grain.   The expansion
is based on  the formal exact solution of the Richardson model. We use this expansion to calculate  finite size corrections to the ground state energy, Matveev-Larkin parameter, and excitation energies.
\end{abstract}

\maketitle    

\section{Introduction}

Since mid 1990's, when Ralph, Black, and Tinkham succeeded in resolving
the discrete
excitation spectrum of nanoscale superconducting metallic grains \cite{RBT}, there has been considerable
 effort  to describe theoretically superconducting correlations in such grains (see e.g. Ref.~\onlinecite{delft} for
a review).  A key question in any such description is how results of the BCS theory are modified in finite systems. 
In this paper we address this problem by developing a systematic expansion in the inverse number of electrons on the grain for the low energy spectrum of the problem.

 In the absence of spin-orbit interaction and magnetic fields one can describe \cite{kaa} superconducting correlations in weakly disordered grains by a simple pairing (BCS) Hamiltonian
\beg
H_{BCS}={\sum_{i,\sigma}\eps_i c_{i \sigma }^\dagger c_{i \sigma}-\lam d\sum_{i,
j=1}^{n} c_{i\dn}^\dagger c_{i\up}^\dagger c_{j\up} c_{j\dn}}
\label{BCS}
\en
Here $\eps_i$ are orbital energy levels and $d$ is the mean level spacing, $d=\langle \eps_{i+1}-\eps_i\rangle$.
Operators $c_{i \sigma }^\dagger$  ($c_{i \sigma }$) create (annihilate) an electron of spin projection $\sigma$ in orbital state $i$, $n$  is the total number of levels, and $\lam$ denotes a dimensionless  coupling constant. The interaction part of Hamiltonian \re{BCS} allows only transitions of singlet electron pairs between the orbitals.

BCS Hamiltonian \re{BCS} is known to be integrable \cite{integrability} and
 solvable by Bethe's {\it Ansatz}.
The exact solution \cite{theSolution} yields a complicated set of
coupled polynomial equations (see Eq.~(\ref{rich}) bellow). As a
consequence, very few
 explicit results have been derived and
most studies resorted to numerics  \cite{delft, dukelsky, imry} based on the exact solution. The
purpose of the
present paper is to remedy   this situation in the regime when the level spacing is the smallest energy scale in the problem.

BCS Hamiltonian (\ref{BCS}) was  studied extensively in 1960's
in the context of pair correlations in nuclear matter \cite{mott}. A straightforward but important observation was
that singly occupied orbitals  do not
participate in  pair scattering \cite{soloviev}.  Hence, the labels of these orbitals are good
quantum numbers and their contribution to the total energy is only through the kinetic  term in BCS Hamiltonian (\ref{BCS}). Due to this  ``blocking effect''
 the problem of diagonalizing  Hamiltonian (\ref{BCS}) reduces to  
the subspace of orbitals that are either empty or doubly  occupied -- ``unblocked''  orbitals.
The latter problem turns out to be solvable \cite{theSolution} by Bethe's {\it Ansatz}.
The spectrum is obtained from the following set of algebraic equations for unknown parameters~$E_i$:
\begin{equation}
-\frac{1}{\lam d}+{\sum_{j=1}^m}\lefteqn{\phantom{\sum}}'\frac{1}{E_i-E_j}=\frac{1}{2}\sum_{k=1}^n\frac{1}{E_i-
\epsilon_k}\quad i=1,\dots,m
\label{rich}
\end{equation}
where $m$ is the total number of singlet pairs and $n$ now is the number of unblocked  orbitals~$\epsilon _k$. Bethe's {\it Ansatz} equations  (\ref{rich}) for BCS Hamiltonian (\ref{BCS})
are commonly referred to as Richardson's equations.
Eigenvalues of BCS Hamiltonian (\ref{BCS}) are related to Richardson parameters, $E_i$, via
\begin{equation}
E=2\sum_{i=1}^{m}
E_i+\sum_B \epsilon_B
\label{E}
\end{equation}
where $\sum_{B} \epsilon_{B}$ is a sum over singly occupied -- ``blocked''  orbitals.

In 1977, Richardson used  exact solution (\ref{rich}) to outline \cite{largeN} a method for expanding the low  energy spectrum in powers of the inverse number of pairs, $1/m$.  
Richardson showed that BCS results \cite{BCS} for the energy gap, condensation energy,
excitation spectrum etc.
 are recovered
from  exact solution (\ref{rich}) in the thermodynamical limit. The proper limit is
obtained by taking the number of levels, $n$, to infinity, so that
$nd\to 2D=\mbox{const}$, $\phantom{.}m=n/2$,
where $D$ is an ultraviolet cutoff usually identified with Debye energy.
In particular, for  equally spaced levels $\eps_i$, the energy gap
$\Delta$ and the ground state energy in the thermodynamical limit are
\beg
\Delta_0(\lambda)=\frac{D}{\sinh(1/\lambda)}\qquad
E_{g.s.}^{BCS}(\lambda)=-Dm\coth{1/\lambda}
\label{gapgr}
\en

In the present paper we show that the ground state and excitation energies of BCS Hamiltonian \re{BCS} can be evaluated explicitly to any order in $d/\Delta_0\sim 1/m$ in terms of the BCS gap  $\Delta_0$, chemical potential $\mu$, mean level spacing $d$, ultraviolet cutoff $D$, and the thermodynamic density of states $\nu(\eps)$. In the physical limit $\Delta_0/D\to 0$, the expansion is applicable  for $\Delta_0\geq d$. In fact, we believe  that in this limit the expansion is in powers of $d/\Delta_0$ with a convergence radius $d/\Delta_0\sim 1$.

BCS Hamiltonian \re{BCS} supports two types of low energy excitations. Excitations of the first type preserve the number of pairs (pair-preserving excitations). The second type of low lying excitations (pair-breaking excitations) is obtained by breaking a single electron pair. In the thermodynamical limit both types of excitations are gapped with the same gap, $\Delta^p=\Delta^b=2\Delta_0$, where $\Delta^p$ and $\Delta^b$ are the energy gaps for pair-preserving and pair-breaking excitations respectively. In Section~\ref{excitation}, we evaluate leading finite size corrections (of order $1/m$) to the gaps $\Delta^p$ and $\Delta^b$. Interestingly, it turns out that these corrections coincide, even though the two gaps are not identical in higher orders in $1/m$. In the limit $\Delta_0/D\to 0$, our result yields $\Delta^p=\Delta^b=2\Delta_0-d$. We also show that the energy levels of lowest excitations of two types cross at certain value of the coupling constant $\lam$.

Another measure of the low energy properties of BCS model \re{BCS} is the parity parameter\cite{ML} introduced by Matveev and Larkin. This parameter is defined as 
\beg
\Delta_{ML}=E_{g.s.}^{2m+1}-\frac{1}{2}\left(E_{g.s.}^{2m+2}+E_{g.s.}^{2m}\right)
\label{ml1}
\en
where $E_{g.s.}^{l}$ is the ground state energy of BCS Hamiltonian \re{BCS} with $l$ electrons. Matveev and Larkin evaluated $\Delta_{ML}$ in the physical limit $\Delta_0/D\to 0$ in two different regimes: $\Delta_0\gg d$ and $\Delta_0\ll d$. They found that in the first regime the leading finite size correction to the parity parameter \re{ml1} comes entirely from the stationary point (mean field) expression for the ground state energy of BCS Hamiltonian \re{BCS}. Here we use our method to calculate $\Delta_{ML}$ in the regime $\Delta_0>d$ for an arbitrary ratio $\Delta_0/D$.  We show that the contribution of quantum fluctuations to the leading finite size correction to $\Delta_{ML}$ behaves as $(\Delta_0/D)\ln(\Delta_0/D)$ for small $\Delta_0/D$.

The ground state energy of pairing  Hamiltonian  \re{BCS} has been discussed recently in a number of papers.  Numerical fits for finite size corrections to the ground state energy in the weak coupling regime, $\lam\ll 1$, have been proposed \cite{imry, dukelsky}. Here we evaluate the leading finite size correction   exactly and find a complete agreement with numerical results \cite{imry, dukelsky} in the weak coupling regime. 

In Ref.~\onlinecite{imry},  authors studied the condensation energy, defined as the difference between the ground state energy  and the expectation value of BCS Hamiltonian \re{BCS} in the Fermi ground state. This difference was calculated in second order perturbation theory in $\lam$ and compared to BCS expression $E_{g.s.}^{BCS}(\lambda)-E_{g.s.}^{BCS}(0)$. The authors found that the two expressions become of the same order when 
$d\leq \Delta_0 \leq \sqrt{Dd}$ and interpreted this as a new, "intermediate", regime of pairing correlations in metallic grains. We argue bellow that, although the finite size correction to the condensation energy indeed becomes of the same order as the BCS result for $d\leq\Delta_0 \leq \sqrt{Dd}$, this fact does not indicate a new physical regime, but is rather an artifact of the model. Main contribution to the finite size correction to the condensation energy comes from energies close to the ultraviolet cutoff $D$ and therefore is beyond limits of applicability of BCS Hamiltonian \re{BCS}. Effects coming from this range of energies can be properly accounted for \cite{chubukov} within the Eliashberg theory \cite{eli}.

The paper is organized as follows. Section~\ref{review} is devoted to the review of a general method \cite{largeN} of $1/m$ expansion  due to Richardson. In Section~\ref{groundstate}, we show that Richardson's results can be used to evaluate ground state and excitation energies of BCS Hamiltonian~\re{BCS} to any order in $1/m$ and explicitly calculate the leading correction to the ground state energy. In Section~\ref{comparison}, we discuss various limits of our results and make a comparison with previous work. Results for the excitation spectrum and Matveev Larkin parameter are collected in 
Sections~\ref{excitation} and \ref{matveev} respectively, where we also determine the gaps for pair-breaking and pair-preserving excitations and discuss the range of applicability of the $1/m$ expansion.

\section{Review of Richardson's $1/m$ expansion}
\label{review}

Here we briefly review Richardson's  $1/m$ expansion \cite{largeN} for the ground state and excitation energies of pairing Hamiltonian \re{BCS}. The details can be found in the original work \cite{largeN}. In subsequent sections we will use Richardson's results to explicitly evaluate finite size corrections to the low energy spectrum of BCS Hamiltonian \re{BCS}. 

Richardson's $1/m$ expansion is based on an electrostatic analogy  to equations
\re{rich}. In this analogy, the roots $E_i$ of equations \re{rich} are interpreted as locations of $m$ two-dimensional free charges of unit strength in the complex plane.  The free charges are subject to a uniform external field $-1/(\lam d)$ and the field of $n$ fixed charges of strength $1/2$ located at the points $\eps_k$ on the real axis. 
The total electrostatic field at a point $z$ associated with the charge distribution   is given by
\begin{equation}
F(z)=\sum_{i=1}^m \frac{1}{z-E_i}-\frac{1}{2}\sum_{k=1}^n \frac{1}{z-\eps_k}-\frac{1}{\lam d}
\label{field}
\end{equation}

The field $F(z)$ contains complete information about the spectrum of BCS Hamiltonian \re{BCS}. For example, the energy spectrum is related to the quadrupole momentum of $F(z)$. Indeed, defining multipole moments  of $F(z)$  by
\begin{equation}
F(z)=\sum_{m=0}^\infty F^{(m)} z^{-m}
\label{multi}
\end{equation}
and expanding equation \re{field} in $1/z$, we obtain
\begin{equation}
E=2\sum_{i=1}^{m} E_i=2 F^{(2)} +\sum_{k=1}^n \eps_k
\label{quadru}
\end{equation}
\begin{equation}
-\frac{1}{\lam d}=F^{(0)}
\label{mono}
\end{equation}
\begin{equation}
m-\frac{1}{2}=F^{(1)}
\label{dipole}
\end{equation}
The $1/m$ expansion is facilitated by the following field equation that can be derived from equations \re{rich} and \re{field}:
\begin{equation}
\frac{d F}{d z}+F^2=\frac{1}{2}\sum_k \frac{1}{(z-\eps_k)^2}+\frac{1}{4}\left(\sum_k \frac{1}{z-\eps_k}+\frac{2}{\lam d}\right)^2-\sum_k \frac{H_k}{z-\eps_k}
\label{feqn}
\end{equation}
where $H_k$ is the field at the location of the fixed charge $\eps_k$ due to the free charges
\begin{equation}
H_k=\sum_i \frac{2}{\eps_k- E_i}
\label{hk}
\end{equation}
Equation \re{feqn} can be solved by expanding the field $F(z)$ in powers of $1/m$ 
\begin{equation}
F(z)=\sum_{r=0}^\infty F_r(z)
\label{fexpn}
\end{equation}
where $F_r(z)$ is of order $m^{1-r}$. It turns out \cite{largeN} that  the lowest order in \re{fexpn}, $F_0(z)$, together with field equation \re{feqn} completely determine the field $F(z)$ to higher orders in $1/m$. Moreover, to obtain higher orders,
 $F_r(z)$ for $r \ge 1$, from $F_0(z)$ one needs to solve only algebraic equations.  
 
 Different
 states of the system are described by different $F_0(z)$. For example, one can show that the
 BCS ground state corresponds to
 \begin{equation}
F_0(z)=-  \sum_k \frac{\sqrt{(z-\mu)^2+\Delta^2} }{2(z-\eps_k) \sqrt{(\eps_k-\mu)^2+\Delta^2}}
 \label{f0}
\end{equation}
The parameters  $\Delta$ and $\mu$ correspond to the BCS gap and chemical potential respectively. Equations for  $\Delta$ and $\mu$ can be derived by  substituting $F_0(z)$ into equations \re{mono} and \re{dipole}
\beg
\frac{2}{\lam d}=\sum_k\frac{1}{\sqrt{(\eps_k-\mu)^2+\Delta^2}}
\label{gap}
\en
\beg
n-2m=\sum_k\frac{\eps_k-\mu}{\sqrt{(\eps_k-\mu)^2+\Delta^2}}
\label{pot}
\en
There are no higher order corrections to equations \re{gap} and \re{pot}, since by construction $F_0(z)$ yields exact monopole and dipole moments of $F(z)$, $F^{(0)}(z)$ and $F^{(1)}(z)$.

Note that, according to equations \re{field} and \re{f0}, $F_0(z)$ also describes the fixed charges  exactly, since
\beg
\lim_{z\to \eps_k} (z-\eps_k) F_0(z)=-\frac{1}{2}
\label{lim}
\en
Higher order corrections to the field $F(z)$ can be expressed only in terms of $\eps_k$, $\Delta$, $\mu$ and finite zeroes of $F_0(z)$
\begin{equation}
 \sum_{k=1}^n \frac{1}{(x_l-\eps_k)\sqrt{(\eps_k-\mu)^2+\Delta^2}}=0
\label{roots}
\end{equation}
For example,
\begin{equation}
F_1(z)=\frac{1}{2Z(z)}\left(\sum_k \frac{z+\eps_k-2\mu}{Z(z)+Z(\eps_k) }-\sum_l \frac{z+x_l-2\mu}{Z(z)+Z(x_l) } -\frac{z-\mu}{Z(z)}\right) 
\label{f1}
\end{equation}
where
$$
Z(z)=\Z{z}
$$
One can show (by e.g. sketching the LHS of equation~\re{roots})  that there are $n-1$ finite solutions to equation \re{roots}, each of them lying between two consecutive single electron levels~$\eps_k$.

The ground state energy to the first two orders in $1/m$, i.e. to the order $m^0$, can be obtained
from $F_0(z)$ and $F_1(z)$ using equation \re{quadru}.
$$
E=E_0+E_1 
$$
\beg
E_0=\sum_k \eps_k -\mu(n-2m)+\frac{\Delta^2}{\lam d}-\sum_k  \Z{\eps_k}
\label{e0}
\en
\beg
E_1=-m\lam d+\sum_{l=1}^{n-1}\left[\sqrt{(x_l-\mu)^2+\Delta^2}-\frac{N_l}{P_l}\right]
\label{e1}
\en
where
$$
N_l=\sum_k\frac{1}{(x_l-\eps_k)^2}\quad P_l=\sum_k
\frac{1}{(x_l-\eps_k)^2\sqrt{(\eps_k-\mu)^2+\Delta^2}}
$$
To calculate  excitation energies one needs to appropriately modify $F_0(z)$, the lowest order in $1/m$ of the electrostatic field $F(z)$.  
Here we  simply write down excitation energies to the first two nonzero orders in $1/m$ referring  readers interested in the detailed derivation to the original work.\cite{largeN}  
$$
e(l)=e_1(l)+e_2(l)\qquad l=1,\dots, n-1
$$
\beg
e_1(l)=2\Z{x_l}
\label{ex1}
\en
\beg
e_2(l)=2\sum_{m\ne l}\frac{1}{P_l}\left[ (F'_1)^2-(F_1)^2+\frac{d}{d z}(F'_1-F_1)+\frac{2F'_1}{x_m-x_l}\right]_{z=x_m}
\label{ex2}
\en
where
\beg
F'_1(z)=F_1(z)+\frac{\Z{x_l}}{(z-x_l)\Z{z}}-\frac{1}{z-x_l}
\label{f'}
\en
and $e(l)$ is the excitation energy relative to the ground state.

Finally, we note that the lowest nonzero order of $1/m$ expansion, $E_0$ and $e_1(l)$ for the ground state and excitation energies, reproduces the mean field (BCS) results for pairing Hamiltonian \re{BCS}. Therefore, the mean field for pairing Hamiltonian \re{BCS} is exact in the thermodynamical limit, while contributions $E_1$ and $e_2(l)$, equations \re{e1} and \re{ex2}, are leading finite size corrections to the thermodynamical limit.

\section{Ground state energy}
\label{groundstate}

Here we evaluate the leading finite size correction to the ground state energy of BCS Hamiltonian \re{BCS}. 

First, we note that, as shown in  Appendix A, expression \re{e1} for the finite size correction $E_1$ can be cast into a simpler form
\beg
E_1=\lambda d\left(\frac{n}{2}-m\right)+\sum_{l=1}^{n-1}\sqrt{(x_l-\mu)^2+\Delta^2}-
\sum_{k=1}^n\sqrt{(\eps_k-\mu)^2+\Delta^2}
\label{koka}
\en
To facilitate comparison to the mean field BCS result \re{gapgr}, we  assume  bellow $n=2m$ equally spaced single electron levels $\eps_k=(k-m-1/2)d$ with energies ranging from $D=(m-1/2)d$ to $-D$. It should be emphasized, however, that explicit results in terms of $\Delta$, $\mu$, and the density of states $\nu(\eps)$ can be equally well obtained  for arbitrary continuous $\nu(\eps)$.

Since $n=2m$ and $\eps_k$ are distributed symmetrically with respect to zero, equation \re{pot} yields $\mu=0$, while equations \re{gap}, \re{e0}, and \re{koka} become
\beg
\frac{2}{\lam d}=\sum_{k=1}^{2m}\frac{1}{\sqrt{\eps_k^2+\Delta^2}}
\label{gap1}
\en
\beg
E_0= \frac{\Delta^2}{\lam d}-\sum_{k=1}^{2m}  \sqrt{\eps_k^2+\Delta^2}
\label{e01}
\en
\beg
E_1= \sum_{l=1}^{2m-1}\sqrt{x_l^2+\Delta^2}-
\sum_{k=1}^{2m}\sqrt{\eps_k^2+\Delta^2}
\label{koka1}
\en
Equation \re{roots} for $x_l$ now reads
\begin{equation}
 f(x_l)=\sum_{k=1}^{2m} \frac{1}{(x_l-\eps_k)\sqrt{\eps_k^2+\Delta^2}}=0
\label{roots1}
\end{equation}
Since for each $\eps_k$ there is $\eps_{k'}=-\eps_k$, $f(z)$ is an odd function of $z$. Therefore, $x_l=0$ is a solution of equation \re{roots1}, while the remaining $n-2=2m-2$ nonzero solutions come in pairs of $x_l$ and $-x_l$. Let us label $m-1$ positive roots $x_l$ with $l=1,\dots, m-1$ and relabel $m$ positive single electron energies $\eps_k$ with $k=0, 1\dots, m-1$. Then, we can rewrite equation~\re{koka1} as
\beg
E_1=\Delta-2\sqrt{\frac{d^2}{4}+\Delta^2}+2\sum_{l=1}^{m-1}\sqrt{x_l^2+\Delta^2}-
\sum_{k=1}^{m-1}\sqrt{\eps_k^2+\Delta^2}
\label{e1int}
\en
where we have separated contributions to the summations of $x_l=0$ and $\eps_k=\pm d/2$. 

Because $x_l$ is located between $\eps_l$ and $\eps_{l-1}=\eps_l-d$, we can write it as
$x_l=\eps_l-\alpha_l d$, where $0<\alpha_l< 1$. Expanding $\sqrt{x_l^2+\Delta^2}$
 in $x_l$ in the vicinity of $x_l=\eps_l$ and bearing in mind that $d\approx D/m$ is of order $1/m$, we obtain
\beg
E_1=-\Delta-2\sum_{l=1}^{m-1}\frac{\alpha_l d}{\sqrt{\eps_l^2+\Delta^2}}
\label{e1semifinal}
\en
where we neglected terms of order $1/m$. With the same accuracy, we can replace the summation over $k$ with an integration
\beg
E_1=-\Delta-2\int_{0}^{D}d\eps \frac{\eps \alpha(\eps)}{\sqrt{\eps^2+\Delta^2}}
\label{e1final}
\en
Note that $E_1$ is indeed of order $m^0$ as it should be. The function $\alpha(\eps)$ is evaluated in Appendix B. The result, up to terms of order $1/m$, is
\beg
\alpha(\eps)=-\frac{1}{\pi}{\rm arccot} \frac{1}{\pi}\ln\left[\frac{D\sqrt{\eps^2+\Delta^2}-\eps\sqrt{D^2+\Delta^2} }{D\sqrt{\eps^2+\Delta^2}+\eps\sqrt{D^2+\Delta^2} }\right]
\label{alpha}
\en
Introducing a new variable
\beg
x= \frac{1}{\pi}\ln\left[\frac{D\sqrt{\eps^2+\Delta^2}-\eps\sqrt{D^2+\Delta^2} }{D\sqrt{\eps^2+\Delta^2}+\eps\sqrt{D^2+\Delta^2} }\right]\qquad
\eps=-\frac{D\Delta \sinh(\pi x/2)}{\sqrt{\Delta^2\cosh^2 (\pi x/2)+D^2}},
\label{change}
\en
we can cast expression \re{e1final} into a more convenient form
\beg
\label{e1ans}
E_1=-2\int_0^\infty \frac{d x}{\pi}\frac{\Delta \sqrt{\Delta^2+D^2}}
{(1+x^2)\sqrt{\Delta^2+D^2 \left( \cosh(\pi x/2)\right)^{-2} } }
\en
To complete the evaluation of the ground state energy to order $m^0$, we also need to calculate the leading term, $E_0$ with the same accuracy. The first step is to replace summation in equations \re{gap} and \re{e01} with integrations according to the following formula:
$$
d\sum_{j=n_1}^{n_2}f(jd)= \int_{n_1d}^{n_2d}dx f(x)+
\frac{d}{2}\left[f(n_1d)+f(n_2d)\right]+o(1/m)
$$
Equations \re{gap} and \re{e01} now read
\begin{equation}
\label{gapint}
\frac{2}{\lam}=\int_{-D}^D \frac{d\eps}{\sqrt{\eps^2+\Delta^2}}+\frac{d}{\sqrt{\Delta^2+D^2}}
\end{equation}
\begin{equation}
\label{e0int}
E_0=\frac{\Delta^2}{\lam d}-\frac{1}{d}\int_{-D}^{D}d\eps \sqrt{\eps^2+\Delta^2}-\sqrt{\Delta^2+D^2} 
\end{equation}
The solution of equation \re{gapint} for $\Delta$ to order $m^0$ is obtained by dropping the second term on the RHS. Evaluating the integral, we obtain $\Delta_0=D/[\sinh(1/\lam)]$ in agreement with equation \re{gapgr}. To compute the correction of order $1/m$ to $\Delta$, we substitute $\Delta=\Delta_0+\delta\Delta$ into equation \re{gapint} and expand in $\delta\Delta$. Keeping only terms of order $1/m$, we find
\beg
\label{newgap}
\Delta=\Delta_0+d \frac{\Delta_0}{2D}
\en
Plugging $\Delta$ into equation \re{e0int} and using $\sqrt{\Delta_0^2+D^2}=D\coth(1/\lam)$, we obtain up to terms of order $1/m$
\beg
\label{e0final}
E_0=-\left(m+\frac{1}{2}\right)D\coth(1/\lam)
\en
Note also that  $\Delta$ in expression \re{e1ans} for $E_1$ can be replaced by $\Delta_0$ up to terms of order $1/m$. Thus, the ground state energy of BCS Hamiltonian \re{BCS} for $m$ pairs and $n=2m$ equally spaced levels is
\beg
E_{g.s.}=-D\coth(1/\lam)\left[m+\frac{1}{2}+\phi(\lam)\right]
\label{ground}
\en
where
\beg
\phi(\lam)=2 \int_0^\infty \frac{d x}{\pi(1+x^2)}\frac{\cosh(\pi x/2)}
{\sqrt{\cosh^2(\pi x/2)+ \sinh^2(1/\lam) } }
\label{phi}
\en
The plot of function $\phi(\lam)$ is shown on Fig.~1.
 \begin{figure}[ht]
\begin{center}
\setlength{\unitlength}{10cm}
\begin{picture}(1, 0.718)(0,0)
   \put(0,0){\resizebox{1\unitlength}{!}{\includegraphics{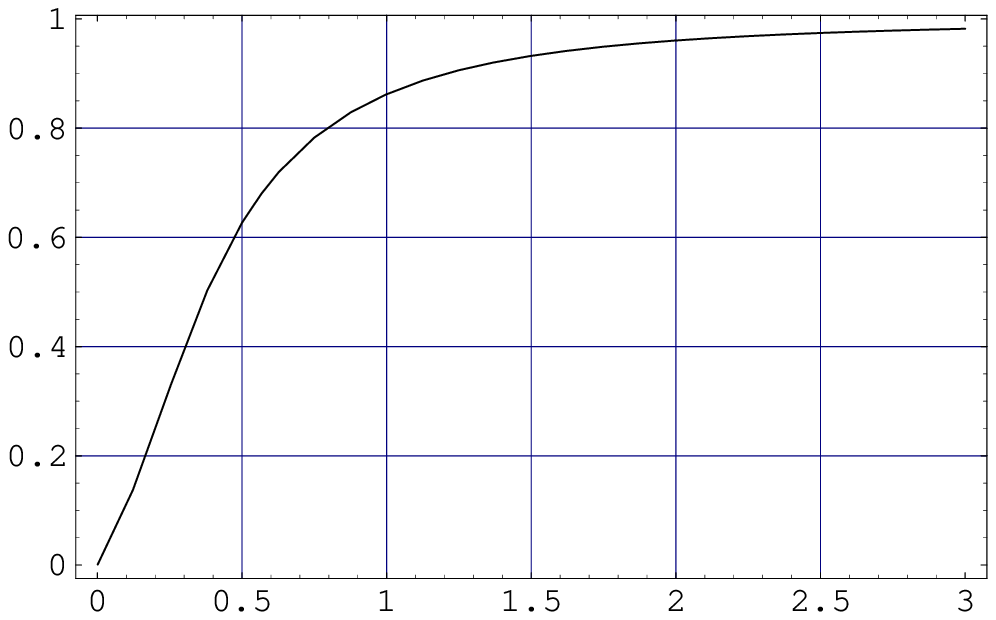}}}
    \put(0.54,-0.03){\makebox(0,0)[b]{\large$\lam$}}
   \put(-0.05,0.35){\rotatebox{90}{\makebox(0,0)[t]{\large $\phi(\lam)$}}}
\end{picture}
\end{center}
\caption{The plot of function $\phi(\lam)$ defined by equation \re{phi}. This function appears in leading finite size corrections  to ground state \re{ground} and excitation (\ref{dbfinal}, \ref{iden1}) energies of BCS Hamiltonian \re{BCS} and to Matveev-Larkin parameter \re{id2}. Note the asymptotics  $\phi(\lam)\to 0$ and $\phi(\lam)\to 1$ for $\lam\to 0$ and $\lam\to\infty$ respectively.}
\end{figure}

The finite size correction to the mean field BCS result \re{gapgr} is
\beg
E_{g.s.}=E^{BCS}_{g.s.}+E_{f.s.}\qquad E_{f.s.}=-D\coth(1/\lam)\left[\frac{1}{2}+ \phi(\lam)\right]
\label{fs}
\en
Note that $E_{f.s.}$ is different from $E_1$ given by equation \re{e1ans} due to contribution of
order $m^0$ from $E_0$.

Higher order corrections to the ground state energy can also be evaluated explicitly. The first step is to express them in terms of $\Delta$ and $x_l$ following prescriptions of Ref.~\onlinecite{largeN}.
Then, $\Delta$ and $x_l$ have to be calculated to appropriate order in $1/m$ using methods of this section and Appendix B. Final results for higher order corrections will involve multiple integrations similar to the integration in equation \re{phi}. For example, the expression for the correction of order $1/m$ contains a triple integral. 

The general case when the distribution of single electron levels in the limit $m, n\to \infty$, $m/n=\mbox{fixed}$ is described by a  continuous density of states $\nu(\eps)$ can be treated similarly. Final expressions for corrections will now be in terms of $\Delta$, $\mu$, and $\nu(\eps)$. For example, the correction of order $m^0$ will be again given by the integral in
equation \re{e1final} where the limits of integration should now be $-D$ and $D$, $\eps$ has to be replaced with $\eps-\mu$, and the integrand has to be multiplied by $\nu(\eps)$ . The function $\alpha(\eps)$ will still be given by equation \re{cot} where now $\nu(\eps)$ has to be included under the integral.

\section{Comparison to previous studies}
\label{comparison}

Here we analyze  our result and compare it to previous results. First, we check whether equation \re{fs} reproduces the results of $1/\lam$ expansion \cite{ourselves} around $\lam=\infty$. Expanding the integrand in equation \re{phi} in $1/\lam$, evaluating the resulting integrals, and expanding $\coth(1/\lam)$  in $1/\lam$, we obtain
$$
E_{f.s.}=-D\left[\frac{3}{2}\lam+\frac{1}{3\lam}-\frac{19}{360\lam^2}+\frac{143}{15120\lam^5}+
O\left(\frac{1}{\lam^7}\right)\right]
$$
Comparing this expression with terms of order $m^0$ in $1/\lam$ expansion  \cite{ourselves} for the ground state energy (see equation (30) of Ref.~\onlinecite{ourselves}), we find that the two results coincide. 

Now let us consider the limit of small $\lam$. The asymptotic behavior of $\phi(\lam)$ for small $\lam$ is worked out in Appendix C. Here we write down the first two terms
\beg
  \phi(\lam)=\lam +\ln 2\cdot\lam^2+O(\lam^3)
 \label{lam}
 \en
 Expanding $\coth(1/\lam)=\sqrt{1+\Delta_0^2/D^2}$ in $\Delta_0/D$ and using $D=(m-1/2)d$, we obtain from equation \re{ground}
\beg
E_{g.s.}=-D\left(m+\frac{1}{2}\right)-\frac{\Delta_0^2}{2d}-D\lam-\ln2 \cdot D\lam^2+O(\lam^3)
\label{imry1}
\en
The first term in equation \re{imry1} is the energy of noninteracting Fermi ground state to order $m^0$. The second term is the nonperturbative mean field (BCS) contribution to the ground state energy. The first two terms are extensive and survive  the thermodynamical limit. The last two terms give the correction to the ground state energy that one would obtain in the  second order of ordinary perturbation theory in $\lam$ around noninteracting Fermi ground state. 

We see that our result \re{fs} yields the leading finite size correction to the thermodynamical limit for all values of $\lam$. In particular, there is no breakdown in the regime of ultrasmall grains, i.e. for $d>\Delta_0$. As we will see in subsequent sections, this is not a generic feature of our approach, but is specific to the ground state energy and is probably related to the ultraviolet nature (see bellow) of the finite size correction calculated above. 

A frequently discussed quantity \cite{delft, dukelsky, imry} is the difference between the  ground state energy and the expectation value of BCS Hamiltonian \re{BCS} in the unperturbed Fermi ground state, $|F.g.s.\rangle$, i.e. a state where   single particle levels bellow the Fermi level, $\eps_k<0$, are doubly occupied, while the remaining levels are empty. This difference is often called condensation energy, even though this name is misleading  for the reasons detailed bellow. However, to facilitate a comparison with results of Ref.~\onlinecite{dukelsky} and  \onlinecite{imry}, we will use the same terminology in this section. We have
$$
E_{cond.}=\langle F.g.s.|H_{BCS}|F.g.s.\rangle-E_{g.s.}=-D\left(m+\frac{1}{2}\right)-2\lam m d-E_{g.s.}
$$
Using $D=(m-1/2)d$ and equation \re{imry1}, we obtain
\beg
E_{cond.}=\frac{\Delta_0^2}{2d}+\ln2 \cdot D\lam^2+O(\lam^3)
\label{imry}
\en
Comparison shows that  the exact result \re{imry} for $E_{cond.}$ to order $m^0$ is in complete agreement with  fits to numerical data.\cite{imry, dukelsky}

Finally, note that the second term in expression \re{imry} is ultraviolet divergent, since it depends explicitly on the ultraviolet cutoff $D$. For pairing by phonons the ultraviolet cutoff $D$ can be identified with the Debye energy $\omega_D$. To properly take into account any effect that comes from energies of the order of $\omega_D$, one needs to go beyond the BCS theory which is  appropriate  only at  energies much lower than $\omega_D$. The contribution from these energies to finite size corrections can be adequately treated \cite{chubukov} within the Eliashberg theory \cite{eli}.   In particular, the hard cutoff at $D=\omega_D$ has to be replaced by a soft effective cutoff due to the $1/\omega^2$ decay of the phonon propagator for frequencies $\omega \gg \omega_D$. 
Therefore, even though the contribution of the finite size correction in equation \re{imry} becomes important for $\Delta_0 \le \sqrt{D d}$, the conclusion of Ref.~\onlinecite{imry} that this is an indication of any new physical regime is not justified.

\section{Excitation energies}
\label{excitation}

In this section we evaluate leading finite size corrections to lowest  excitation energies. As we will see bellow, the results of this section are accurate only in the regime of relatively large grains, $\Delta_0 > d$, i.e. within terms of order $o(d/\Delta_0)$. These higher order corrections can also be straightforwardly calculated  using methods of Sections~\ref{groundstate}. However, we will only evaluate corrections of order $d/\Delta_0$ here.

As in Section~\ref{groundstate}, we will perform calculations for the case of $2m$ electrons and $n=2m$ equally spaced levels  $\eps_k=(k-m-1/2)d$ with energies ranging from $D=(m-1/2)d$ to $-D$. 
In this case, equation \re{pot} implies $\mu=0$.
A more general case when the single electron levels are distributed with a smooth density of states can be treated similarly (see the discussion bellow equation \re{fs}).

Note that Hamiltonian \re{BCS} conserves the number of paired electrons. Therefore, the excitations can be grouped into two types: those that preserve the number of pairs
and those that break pairs. Energies of low lying pair-preserving excitations in the thermodynamical limit are given by equation \re{ex1} with $\mu=0$
\beg
e^{p}_1=2\sqrt{x_l^2+\Delta_0^2}
\label{ex1re}
\en
where $x_l$ are the roots of equation \re{roots}. Low lying pair-breaking excitations  are obtained by breaking a single pair and placing the two unpaired electrons on two single electron levels $\eps_a$ and $\eps_b$. The energy of this excitation according to equation \re{E} is
\beg
e^b=\eps_a+\eps_b+E_{g.s.}(\eps_a, \eps_b)-E_{g.s.}
\label{break}
\en
where $E_{g.s.}(\eps_a, \eps_b)$ is the ground state energy of BCS Hamiltonian \re{BCS} with levels $\eps_a$ and $\eps_b$ blocked. In the thermodynamical limit, using equation \re{e0}, we obtain 
\beg
e^b_1=\sqrt{\eps_a^2+\Delta_0^2}+\sqrt{\eps_b^2+\Delta_0^2}
\label{br1}
\en
Therefore, in the thermodynamical limit both types of excitations are gapped with the same gap $2\Delta_0$, i.e.
\beg
\Delta^p_1=\Delta^b_1=2\Delta_0
\label{gaps}
\en
Since pair-breaking excitations are capable of carrying spin-1, $\Delta^b$ can also be called 
the spin gap.
To calculate corrections to $\Delta^p_1$ and $\Delta^b_1$, one needs to go beyond mean field approximation.

First, let us determine the energy of lowest lying pair-breaking excitations to order $1/m$. 
Breaking a pair changes both the number of pairs to $m'=m-1$ and also the number of unblocked levels to $n'=2m-2=2m'$.  The lowest energy is archived by  blocking levels $\eps_a=d/2$ and $\eps_b=-d/2$. Since  this leaves the distribution of single particle levels symmetric with respect to zero, the chemical potential $\mu$ in equation \re{pot} remains equal to zero, $\mu'=\mu=0$. However, the blocking affects  the gap $\Delta'$, since now terms corresponding to $\eps_k=\pm d/2$ have to be excluded from gap equation \re{gap}. Using equation \re{gap}, we obtain
\beg
\sum_k\frac{1}{\sqrt{ \eps_k^2+\Delta'^2 } }=\frac{2}{\sqrt{d^2/4+\Delta^2}}+\sum_k\frac{1}{\sqrt{\eps_k^2+\Delta^2}}
\label{dif}
\en
where $\Delta'$ is the value of the gap with levels $\pm d/2$ blocked. Expanding the LHS of equation \re{dif} in $\delta\Delta=\Delta'-\Delta$ and using gap equation \re{gap}, we obtain
\beg
\delta\Delta=-d\sqrt{1+\frac{\Delta^2}{D^2} }
\label{dif1}
\en
According to equation \re{break}, to order $1/m$ the lowest lying pair-breaking excitations
have the following energy:
\beg
\Delta^b=E'_0(\Delta')-E_0(\Delta)+E'_1(\Delta', x'_l)-E_1(\Delta, x_l)
\label{db}
\en
where $E_0(\Delta)$ and $E_1(\Delta, x_l)$ are given by equations \re{e01} and \re{koka1} respectively and primes denote quantities for the ground state with levels $\pm d/2$ blocked.
Equations \re{e01}, \re{koka1}, \re{dif1}, and \re{newgap} imply
\beg
E'_0(\Delta')-E_0(\Delta)=2\Delta'+\sum_k\frac{\delta\Delta \Delta}{4(\eps_k^2+\Delta^2)^{3/2}}=2\Delta_0+\frac{d\Delta_0}{D}-d\sqrt{1+\frac{\Delta_0^2}{D^2}}
\label{e0e0}
\en
\beg
E'_1(\Delta', x'_l)-E_1(\Delta, x_l)=\frac{\partial E_1(\Delta)}{\partial \Delta}\delta\Delta+\sum_l\frac{x_l\delta x_l}{\sqrt{x_l^2+\Delta^2}}
\label{e1e1}
\en
where $E_1(\Delta)$ is given by equation \re{e1ans} and $\delta x_l$ is the change in $x_l$ due to blocking levels $\pm d/2$. 

We see from equation \re{roots1} that the effect of removing levels $\eps_k=\pm d/2$ from the summation in equation \re{roots} is strongest for the roots closest to the blocked levels $\pm d/2$. For these roots $\delta x_l\sim d$. On the other hand, due to an additional factor of $x_l$ in front of $\delta x_l$ in equation \re{e1e1}, the contribution of each of these $x_l$ to the RHS of equation \re{e1e1} is of order~$d^2/\Delta$. By splitting the sum in equation \re{roots1} into two sums as in Appendix B, one can show that the contribution of all these roots to the sum in equation \re{e1e1} is of order $o(1/m)$. For the remaining roots, $\delta x_l/ x_l$ is of order $1/m$ and each term in equation \re{roots1} can be expanded into $\delta x_l/(x_l-\eps_k)$. We have
$$
\sum_{\eps_k\ne \pm d/2}\frac{1}{\sqrt{\eps_k^2+\Delta^2}(x'_l-\eps_k)}=\sum_{k=1 }^{2m}\frac{1}{\sqrt{\eps_k^2+\Delta^2}(x_l+\delta x_l-\eps_k)}-\frac{2}{x_l \Delta}=\sum_{k=1 }^{2m}\frac{1}{\sqrt{\eps_k^2+\Delta^2}(x_l-\eps_k)}
$$
Expanding into $\delta x_l$, we obtain
$$
\delta x_l\sum_k\frac{1}{\sqrt{\eps_k^2+\Delta^2}(x_l-\eps_k)^2}=-\frac{2}{x_l \Delta}
$$
The summation here can be evaluated in the same way as the first sum in equation \re{app1} of Appendix B is evaluated. Recall that roots of equation \re{roots1} $x_l$ and therefore $\delta x_l$ are distributed symmetrically with respect to zero. Using the notation introduced in the text following  equations \re{roots1} and \re{e1int}, we have for $x_l>0$
$$
\delta x_l x_l=-\frac{2d^2\sqrt{\eps_l^2+\Delta^2}}{\Delta}\frac{\sin^2\pi\alpha(\eps_l)}{\pi^2}
$$ 
where $\alpha(\eps_l)$ is given by equation \re{alpha}. Substituting  $\delta x_l x_l$ into equation \re{e1e1} and using equations \re{e0e0}, \re{db}, \re{alpha}, and \re{e1ans}, we obtain
\beg
\Delta^b=2\Delta_0-d\sqrt{1+\frac{\Delta_0^2}{D^2}}+\frac{d\Delta_0}{D}\left[1+\phi(\lam)\right]
\label{dbfinal}
\en
where we used the change of variables \re{change} and $\phi(\lam)$ is defined by equation \re{phi}. Expression \re{dbfinal} yields the energy of lowest lying pair-breaking excitations up to terms of order $o(d/(\min[ D, \Delta_0]))$.

In the physical limit of weak coupling, $\Delta_0/D\to 0$,  according to equation \re{lam}, expression \re{dbfinal} becomes
\beg
 \Delta^b=2\Delta_0-d+o(d/\Delta_0)
 \label{phys1}
 \en
 
Next, we turn to excitations that preserve the number of pairs. Energies of these excitations to order $1/m$ are given by equations \re{ex1} and \re{ex2}.  Equation \re{ex1re} shows that the lowest lying excitation corresponds to $x_l=0$. We have, up to terms of order  $o(d/(\min[ D, \Delta_0]))$
\beg
\Delta^p=2\Delta+2\sum_{x_m\ne 0}\frac{1}{P_l}\left[ (F'_1)^2-(F_1)^2+\frac{d}{d z}(F'_1-F_1)+\frac{2F'_1}{x_m}\right]_{z=x_m}
\label{aux}
\en
where $F_1(z)$ and $F'_1(z)$ are defined by equations \re{f1} and \re{f'}. Taking into account
that both $\eps_k$ and $x_l$ are distributed symmetrically with respect to zero and  $\mu=0$, we can rewrite these equations as 
$$
F_1(z)=\frac{z}{2\sqrt{z^2+\Delta^2}}\left(\sum_k \frac{1}{\sqrt{z^2+\Delta^2}+\sqrt{\eps_k^2+\Delta^2 } }-\sum_l \frac{1}{\sqrt{z^2+\Delta^2}+\sqrt{x_l^2+\Delta^2  }} -\frac{1}{\sqrt{z^2+\Delta^2}}\right) 
$$
$$
F'_1(z)=F_1(z)+\frac{\sqrt{x_l^2+\Delta^2}}{(z-x_l)\sqrt{z^2+\Delta^2}}-\frac{1}{z-x_l}
$$
Summations in $F_1(z)$ and in equation \re{aux} can be evaluated in the same way as sums in
equations \re{e1e1} and  \re{e1int} have been evaluated. Even though this calculation looks rather different from the one that lead to equation \re{dbfinal}, it  yields an identical result, i.e.
\beg
\Delta^p=\Delta^b+o(d/(\min[ D, \Delta_0]))
\label{iden1}
\en
 
Thus, both gaps coincide up to terms of order $o(1/m)$. However, this coincidence is not preserved in higher orders. Indeed, it was shown in Ref.~\onlinecite{ourselves} that in the strong coupling limit, $\lam\gg 1$, the gap for pair-breaking excitations is larger $\Delta^b-\Delta^p\simeq d^2/\Delta_0>0$. On the other hand, at $\lam=0$ the gap for pair-preserving excitations is larger, $\Delta^b-\Delta^p=-d$. Therefore, the lowest energy levels of the two types of excitations cross at certain value of $\Delta_0$. Equation \re{iden1} shows that the distance between the two levels is  reduced from $d$ at $\Delta_0$ to $o(d/\Delta_0)\phantom{.}d$ even when $d\ll \Delta_0\ll D$. However, the knowledge of higher order corrections to the gaps $\Delta^b$ and $\Delta^p$ is needed to determine whether the crossing occurs in the physical regime $\Delta_0/D\to 0$, i.e. at $\Delta_0 \simeq d$.

\section{Matveev-Larkin parameter}
\label{matveev}

Finally, let us evaluate the Matveev-Larkin parameter \cite{ML}. This parameter is a measure of a parity effect in the grain and is defined as follows:
\beg
\Delta_{ML}=E_{g.s.}^{2m+1}-\frac{1}{2}\left(E_{g.s.}^{2m+2}+E_{g.s.}^{2m}\right)
\label{ml}
\en
where $E_{g.s.}^{l}$ is the ground state energy of BCS Hamiltonian \re{BCS} with $l$ electrons.

The calculation of $\Delta_{ML}$ is similar to the one that lead to equation \re{dbfinal}, only now we also have to take into account the change in the chemical potential
$$
\begin{array}{l}
\mu_{2m+2}-\mu_{2m}=2(\mu_{2m+1}-\mu_{2m})=-2(\Delta_{2m+2}-\Delta_{2m})=
d\sqrt{1+\frac{\Delta_0^2}{D^2} }\\
\\
 \Delta_{2m+2}-\Delta_{2m}=O(d^2/\Delta_0)\\
 \end{array}
$$
The calculation results in
\beg
\Delta_{ML}=\frac{\Delta^b}{2}=\Delta_0 -\frac{d}{2}
\sqrt{1+\frac{\Delta_0^2}{D^2}}+\frac{d\Delta_0}{2D}\left[1+\phi(\lam)\right]
\label{id2}
\en
where $\phi(\lam)$ is defined by equation \re{phi}.
As before, this expression is accurate up to terms of order $o(d/(\min[\Delta_0, D]))$. In the physical limit $\Delta_0/D\to 0$, expression \re{id2},  according to equation \re{lam}, reduces to the one obtained in Ref.~\onlinecite{ML}
\beg
\Delta_{ML}=\Delta_0-\frac{d}{2}+o(d/\Delta_0)
\label{phys2}
\en

The first three terms on the RHS of equation \re{id2} come from the mean field (stationary point) approximation \re{e0} for the ground state energy. The last term in equation \re{id2} represents the contribution of order $1/m$ of quantum fluctuations around the stationary point. The asymptotic behavior of this term in the physical limit $\Delta_0/D\to 0$ is given by equation \re{lam}.  In terms of $d$, $\Delta_0$, and $D$ it reads $d\ln (\Delta_0/D)\Delta_0/D$. In this limit quantum fluctuations will contribute to higher orders in $d/\Delta_0$ as evidenced by the result\cite{ML} for $\Delta_{ML}$ 
in the regime $d\ll \Delta_0$. Therefore, it is of certain interest to use methods of Section~\ref{groundstate} to evaluate further corrections to $\Delta_{ML}$.

We conclude this section with a comment on the range of applicability of $1/m$ expansion detailed in this paper. It is clear from equations \re{phys1} and \re{phys2} that the expansion is applicable in the regime
$\Delta_0 \geq d$. In fact,  results of Ref.~\onlinecite{ourselves} and \onlinecite{largeN} (see also Section~\ref{review}) suggest that the expansion is in powers of $d/\Delta_0$ with a convergence radius $d/\Delta_0\simeq 1$.

\section{Conclusion}

In this paper we have shown that finite size corrections to the thermodynamical limit for pairing Hamiltonian \re{BCS} can be evaluated explicitly in terms of the BCS gap $\Delta_0$, chemical potential $\mu$, mean level spacing $d$, ultraviolet cutoff $D$, and the thermodynamic density of states $\nu(\eps)$ to any order in $d/\Delta_0\sim 1/m$. 

We evaluated leading corrections to the ground state and lowest excitation energies, and to Matveev-Larkin parameter (equations (\ref{fs}, \ref{dbfinal}, \ref{phys1}, \ref{iden1}, \ref{id2}, \ref{phys2})). Our results for the ground state energy are in agreement with previous numerical studies.  We showed that the finite size correction to the condensation energy is ultraviolet divergent and therefore comparing it to the BCS result is not justified.

We found that the gaps for pair-breaking and pair-conserving excitations of pairing Hamiltonian \re{BCS} coincide up to terms of order $o(1/m)$, where $m$ is the number of electron pairs on the grain. In higher orders in $1/m$ the two gaps are different, the difference being of order $d^2/\Delta_0$, where $d$ is the mean level spacing and $\Delta_0$ is the BCS gap  \re{gapgr}. We showed that the energy levels of the lowest excitations of two types cross at a certain value of the coupling constant $\lam$.

The range of applicability of $1/m$  expansion detailed in the present paper is $\Delta_0\geq d$. In fact, we believe that in the physical limit $\Delta_0/D\to 0$ the expansion is a power series  in $d/\Delta_0$ with a convergence radius of order one.

Note that our results significantly simplify in the physical limit $\Delta_0/D\to 0$ (e.g. compare equations \re{dbfinal} and \re{phys1}). An interesting open problem is to take this limit directly in Richardson's equations \re{rich} and to develop a simplified version of the $1/m$ expansion for this case. In particular, this might help to address the problem of the crossover  between the fluctuation dominated ($d\gg \Delta_0$) and the bulk  ($d\ll \Delta_0$) regimes.

\section{Acknowledgements}

We are grateful to Akaki Melikidze for showing to us how expression \re{e1} can be simplified to equation \re{koka}. We thank Igor Aleiner for useful discussions. One of the authors, B. L. A.,  also acknowledges the support  of EPSRC under the grant  GR/S29386.  

\section{Appendix A}
\renewcommand{\theequation}{A.\arabic{equation}}
Here we show that expression \re{e1} for the correction to the ground state energy can be simplified to equation \re{koka}. Indeed, define
$$
f(z)= \sum_{k=1}^n \frac{d_k}{z-\eps_k}\qquad \mbox{ where $\dis d_k=\frac{1}{\Z{\eps_k}}$ }
$$
Equation \re{roots} now reads $f(x_l)=0$. The function $f(z)$ has $n-1$ finite zeroes at $z=x_l$ and also a zero at $z=\infty$. Its dual function, $g(z)=1/f(z)$, has $n-1$ poles at $z=x_l$ and also a pole at $z=\infty$ with a residue $(\sum_{k=1}^n d_k)^{-1}$. Therefore, it can be represented as
$$
g(z)=\sum_{l=1}^{n-1} \frac{m_l}{z-x_l}+\frac{z}{\sum_k d_k}=\sum_{l=1}^{n-1} \frac{m_l}{z-x_l}+\frac{\lam d z}{2}
$$
where we have used $\sum_k d_k=2/(\lam d)$ in accordance with gap equation \re{gap}. The following equations for the residues of $g(z)$ and $f(z)$ are helpful:
$$
m_l=\lim_{z\to x_l} [(z-x_l) g(z)]=\lim_{z\to x_l} \frac{z-x_l}{f(z)}=\frac{1}{f'(x_l)}=
-\left[ \sum_k \frac{d_k}{(x_l-\eps_k)^2}\right]^{-1}=-\frac{1}{P_l}
$$
$$
\frac{1}{d_k}=\frac{1}{\lim_{z\to\eps_k} [(z-\eps_k)f(z)]}=g'(\eps_k)=-\sum_l \frac{m_l}{(x_l-\eps_k)^2}+\frac{\lam d}{2}
$$
where the prime denotes the derivative with respect to $z$. Using these equations, we obtain
\beg
\sum_{l=1}^{n-1} \frac{N_l}{P_l}=-\sum_{l=1}^{n-1}\sum_{k=1}^n \frac{m_l}{(x_l-\eps_k)^2}=
\sum_{k=1}^n\left(\frac{1}{d_k}-\frac{\lam d}{2}\right)=-\frac{\lam d n}{2}+\sum_{k=1}^n\Z{\eps_k}
\label{nlpl}
\en
Finally, substituting equation \re{nlpl} into expression \re{e1}, we obtain equation \re{koka}.

\section{Appendix B}
\renewcommand{\theequation}{B.\arabic{equation}}
In this Appendix we solve equation \re{roots1} for $x_l$. As was discussed bellow equation \re{f1}, each solution $x_l$ lies  between two consecutive single electron levels $\eps_k$. Consider the solution $x(\eps)$ that lies  between $\eps-d$ and $\eps$, where we dropped  subscripts for simplicity. 

Now let us multiply  equation \re{roots1} by $d$ and rewrite it as
\beg 
 \sum_{|\eps_k-\eps|\le Jd} \frac{d}{(x(\eps)-\eps_k)\sqrt{\eps_k^2+\Delta^2}}
 +\sum_{|\eps_k-\eps|> Jd} \frac{d}{(x(\eps)-\eps_k)\sqrt{\eps_k^2+\Delta^2}}=0
 \label{app1}
 \en
 where $1\ll J\ll \Lambda=\min[\Delta, D]/d$. For example, one can choose $J=\sqrt{\Lambda}$.
 In the first summation in equation \re{app1}, $\sqrt{\eps_k^2+\Delta^2}$ can be replaced by $\sqrt{\eps^2+\Delta^2}$ with a relative error of order $J d/\Delta$. We obtain
$$
\sum_{|\eps_k-\eps|\le Jd} \frac{d}{(x(\eps)-\eps_k)\sqrt{\eps_k^2+\Delta^2}}=
\left[1+O\left(\frac{Jd}{\Delta}\right)\right] \frac{1}{\sqrt{\eps^2+\Delta^2}} 
\sum_{p=0}^J \left[\frac{1}{p+1-\alpha(\eps)}- \frac{1}{p+\alpha(\eps)} \right]
$$
where $\alpha(\eps)$ is defined by $x(\eps)=\eps-\alpha(\eps) d$. To determine $\alpha(\eps)$ to the leading ($m^0$) order in $1/m$, we can now take the limit $m\to\infty$. With a suitable choice of $J$ (e.g. $J=\sqrt{\Lambda}$), $J\to\infty$ and $(Jd)/D\to 0$ in this limit, while the second sum in equation \re{app1} becomes a principal value integral.   Using,
$$
\sum_{p=0}^\infty \left[ \frac{1}{p+1-\alpha(\eps)}- \frac{1}{p+\alpha(\eps)} \right]=-\pi\cot(\pi \alpha(\eps))
$$
we obtain
\beg
\pi\cot(\pi \alpha(\eps))=\dashint_{-D}^D \frac{d\eps'}{(\eps-\eps')\sqrt{\eps'^2+\Delta^2}}
\label{cot}
\en
Finally, evaluating the integral, we arrive at equation \re{alpha}. 

Corrections $\delta \alpha(\eps)$ to $\alpha(\eps)$ of order $1/m$ and higher can also be evaluated explicitly by expanding  equation \re{roots1} in $\delta \alpha(\eps)$. These corrections contribute to terms of order $1/m$ and higher in the ground state energy.

\section{Appendix C}

Here we determine the asymptotic behavior for small $\lam$ of the integral
$$
\phi(\lam)=2 \int_0^\infty \frac{d x}{\pi(1+x^2)}\frac{\cosh(\pi x/2)}
{\sqrt{\cosh^2(\pi x/2)+ \sinh^2(1/\lam) } }
$$
 First, we note that up to terms of order $e^{-1/\lam}$, one can rewrite this
integral as
$$
\phi(\lam)=2 \int_{-\infty}^\infty \frac{dx}{\pi[1+(x+x_0)^2]\sqrt{1+e^{-\pi x}} }\qquad\mbox{where $\dis x_0=\frac{2}{\pi\lam}$}
$$
Let us divide the domain of integration into three intervals: $(-\infty, -a)$, $(-a, a)$, and
$(a, \infty)$, where $1\ll a\ll x_0$, and denote the corresponding integrals by $I_3$, $I_2$, and $I_1$ respectively. Each of the integrals $I_k$ can be expanded into its own small parameter that depends on $a$. The dependence on $a$ will cancel out when the results are added together. We have
$$
I_3=2\int_a^\infty  \frac{dx}{\pi[1+(x-x_0)^2]\sqrt{1+e^{\pi x}} }=
2\int_a^\infty  \frac{dx}{\pi}\frac{e^{-\pi x/2}-e^{-3\pi x/2}+\dots}{1+(x-x_0)^2}=O(e^{-\pi a/2})
$$
$$
\begin{array}{l}
\dis I_2=2 \int_{-a}^a  \frac{dx}{\pi[1+(x_0+x)^2]\sqrt{1+e^{-\pi x}} }=
\frac{2}{x_0^2}\int_{-a}^a  \frac{dx}{\pi \sqrt{1+e^{-\pi x}} }-\\
\\
\dis \phantom{I_2=}\frac{2}{x_0^3}\int_{-a}^a  \frac{x dx}{\pi \sqrt{1+e^{-\pi x}} }+\dots=\frac{2\pi a+4\ln 2}{\pi^2 x_0^2}+O\left(\frac{a^2}{x_0^3}\right)\\
\end{array}
$$
$$
I_1=2\int_a^\infty  \frac{dx}{\pi[1+(x+x_0)^2]\sqrt{1+e^{\pi x}} }=2\int_a^\infty  \frac{dx}{\pi}\frac{1-e^{-\pi x}/2+\dots}{1+(x+x_0)^2}=\frac{2}{\pi x_0}-\frac{2a}{\pi x_0^2}
+O\left(\frac{a^2}{x_0^3}\right)
$$
Adding $I_1$, $I_2$, and $I_3$, we obtain equation \re{imry1}. Higher order terms can also be calculated by the same method.

\end{document}